\documentclass[11pt]{article}
\usepackage[final]{graphics}
\usepackage{latexsym}

\newcommand{\be}{\begin{equation}}
\newcommand{\ee}{\end{equation}}

\newcommand{\ba}{\begin{eqnarray}}
\newcommand{\ea}{\end{eqnarray}}

\newcommand{\bat}{\begin{tabular}{lr}}
\newcommand{\eat}{\end{tabular}}

\newcommand{\bay}{\[\begin{array}{lr}}
\newcommand{\eay}{\end{array}\]}

\newcommand{\baa}{\begin{eqnarray*}}
\newcommand{\eaa}{\end{eqnarray*}}

\def\lg{{\lambda}}
\def\gam{{\Gamma}}
\def\rs{{r_s}}

\def\br{{\mathbf{r}}}
\def\ror{{\rho(\br)}}
\def\rror{{\left[\ror\right]}}
\def\brp{{\mathbf{r}'}}

\def\rs{{r_s}}

\def\rrp{{|\br}-{\br}'|}

\def\dr{d\br}
\def\drp{d\brp}

\begin{document}
\bibliographystyle{prbsty}
 
\title{The plasma-insulator transition of spin-polarized Hydrogen} 
 
\author{Hong Xu $^{*}$ and Jean-Pierre Hansen $^{\dag}$}
\date{}
\maketitle

\noindent
$^*$ D\'epartement de Physique des Mat\'eriaux (UMR 5586 du CNRS),\\
Universit\'e Claude Bernard-Lyon1, 69622 Villeurbanne Cedex, France\\
$^{\dag}$ Department of Chemistry, University of Cambridge\\
Lensfield Road, Cambridge CB2 1EW, UK\\

\vspace{2truecm}
\pagebreak
 
\begin{abstract}
A mixed classical-quantum density functional theory
is used to calculate pair correlations and the free energy
of a spin-polarized Hydrogen plasma.
A transition to an atomic insulator phase is estimated to
occur around $r_s=2.5$ at $T=10^4K$, and a pressure
$P\approx0.5Mbar$. Spin polarization is imposed to
prevent the formation of $H_2$ molecules.
\end{abstract} 

\pagebreak

%%%%%%%%%%%%%%%%%%%%%%%%%%%%%%%%%%%%%%%%%%%%%%%%%%%%%%%%%%%%%%%%%
\vspace{1truecm}
Although Hydrogen is generally considered to be the simplest
of elements, its expected metallization under pressure~\cite{wig}
has proved a rather elusive transition. It is now accepted
that the behaviour of solid and fluid molecular Hydrogen
($H_2$) may be very different at high pressures. Despite
considerable experimental efforts with static, room temperature
compression of solid $H_2$ in diamond anvils~\cite{che} beyond
pressures of 2 Mbar, there is still no compelling
evidence for a metallic state~\cite{edw}. The situation is somewhat more favourable for fluid
$H_2$, since shock compression to 1.4Mbar, and a temperature of
about 3000K, led to measurements of metallic resistivities~\cite{wei}.
However, theoretical interpretation is hampered by the absence
of a clear-cut scenario; in particular it is not clear whether
molecular dissociation precedes ionization or conversely~\cite{col}.
The presence of several species, $H_2$, $H_2^+$, H, $H^+$ and electrons
at a ``plasma phase transition"~\cite{sau} complicates a
theoretical analysis considerably; to gain a clearer picture
of pressure-induced ionization, it may be instructive to consider a model
system, which would not involve molecular dissociation.\par
The model system considered in this letter is spin-polarized
Hydrogen. If all electron spins are assumed to be polarized
by a strong external magnetic field ${\bf B}$, only triplet
pair states $^3\Sigma$ can be formed, preventing the binding into
$H_2$ molecules. The low pressure phase will be made up
of H atoms and the only possible scenario upon compression will
be the ionization of atoms to form an electron - proton plasma, which
is expected to be crystalline at low temperature and fluid at
higher temperatures. A rough estimate of the magnetic field needed
to spin - polarize the electrons is obtained by equating the
magnetic coupling energy $-\mu_B B$ (where $\mu_B$ is the magnetic
moment of an electron) to the difference between the triplet and
singlet H-H potential energy functions, calculated at the equilibrium
distance of the $H_2$ molecule~\cite{sau}; this leads to 
$B\approx10^5Tesla$. This value exceeds the highest magnetic fields
achievable in a laboratory by 3 orders of magnitude, but is well
within the range of astrophysical situations.
The present calculation neglects possible orbital effects due to a strong 
applied magnetic field; such effects are expected to be small and their
inclusion would lead to a much more involved and less transparent
calculation. We prefer to think of our model system as a plasma which has
been prepared in a spin-polarized state, and is assumed to remain such
even when B is switched off. The subsequent 
calculation will be restricted to fluid Hydrogen.\par 
The thermodynamic properties of the low pressure atomic $H\uparrow$
phase may be easily calculated from the known triplet pair potential~\cite{eak}
by standard methods of the theory of classical fluids. We calculated
the atom-atom pair distribution function $g(r)$ from the HNC integral 
equation~\cite{han}, and deduced from it the equation of state via the
virial and compressibility routes. The resulting excess free energies
per atom are plotted in Fig.4 as a function of the usual density
parameter $r_s=a/a_0$, along the isotherm $T=10^4K$; here $a_0$ is
the Bohr radius, and $a=[3/(4\pi\,n)]^{1/3}$, where $n$ is the number of $H$ atoms
per unit volume. There is a thermodynamic inconsistency, typical of
HNC theory, but the small difference between the ``virial" and
``compressibility" free energies will have no influence on our 
conclusions. To allow for a meaningful comparison with the free
energy calculated for the high pressure plasma phase, the free energies
shown in Fig.4 contain an electron binding energy contribution
of $-0.5\,a.u.$ . It is implicitly assumed that this binding energy,
valid for isolated atoms (i.e. in the limit $\rs\rightarrow\infty$) does not change upon compression up to $r_s=2.5$, due to overlap and distortion of
the individual electron $1s$ orbitals.\par
A statistical description of the high pressure phase is more challenging.
The key parameter characterizing the spin-polarized electron component
is its Fermi energy $\epsilon_F=2.923/r_s^2\;a.u.$ ;the corresponding
Fermi temperature $T_F\approx 9.2\,10^5/r_s^2\;K$. Along the isotherm
$T=10^4K$ considered in the present calculations, the electrons may
be considered to be completely degenerate (i.e. in their ground state)
up to $r_s\approx3$. The degeneracy temperature of the protons is
$2000$ times smaller, so that for $T=10^4K$, the latter may be considered
as being essentially classical, down to $r_s\approx0.5$. The proton
component is characterized by the Coulomb coupling constant
$\gam=e^2/(ak_BT)=31.56/\rs$ along the above isotherm, showing
that classical Coulomb correlations are expected to be strong
over the density range $1\leq \rs\leq3$ considered in this paper.
Note that while $\gam$ decreases as $\rs$ increases, the corresponding
electron Coulomb coupling constant $\gamma=e^2/(a\epsilon_F)=0.342\,\rs$
increases.\par
In the ultra-high density regime $\rs\le1$, the electron kinetic
energy dominates, and the proton and electron components decouple
in first approximation (``two-fluid" model); the weak proton-electron
coupling may be treated by linear response theory~\cite{gal}, suitably
adapted to the spin-polarized case. Within linear response, the free
energy per atom (ion-electron pair) splits into three terms:
the ground-state energy of the uniform, spin-polarized electron
gas (``jellium"), $\epsilon_e$, the free energy of protons in a uniform
neutralizing background (the so-called ``one-component plasma" or OCP),
$f_{OCP}$, and the first order correction due to linear screening
of the Coulomb interactions by the electron gas, $\Delta f$:
\be
f=\frac{F(\gam,\rs)}{N}=\epsilon_e+f_{OCP}+\Delta f
\label{eq1}
\ee
where $\epsilon_e(\rs)$ is taken to be the sum of kinetic ($1.754/\rs^2$),
exchange ($-0.5772/\rs$) and correlation~\cite{par} contributions;
$f_{OCP}$ is given by an accurate fit to Monte Carlo simulations of the
OCP~\cite{gal,sla}; $\Delta f$ follows from first order thermodynamic
perturbation theory~\cite{gal}:
\be
\Delta f=\frac{1}{2(2\pi)^3}\int S_{OCP}(k)\hat{w}(k)d\mathbf{k}
\label{eq2}
\ee
where $S_{OCP}(k)$ is the static structure factor of the OCP (which
plays the role of reference system). According to linear response
theory, $\hat{w}(k)$ is the difference between screened and bare
ion-ion pair potentials:
\be
\hat{w}(k)=\frac{4\pi e^2}{k^2}\left[ \frac{1}{\mathit{\epsilon(k)}
} -1 \right]
\label{eq3}
\ee
where $\mathit{\epsilon(k)}$ is the dielectric function of the electron gas
which we calculated within the RPA from the Lindhard susceptibility
of a gas of spin-polarized, non-interacting electrons, supplemented
by a local exchange and correlation correction~\cite{par}. All necessary ingredients for the calculation
of $\Delta f$ may be found in \cite{gal}, and the resulting free energy curve
is shown in Fig.4. Although linear response cannot, a priori, be
expected to be quantitatively accurate for $\rs>1$, it provides a
rough estimate of the plasma to atomic phase transition,
from the intersection of the free energy curves, which is seen
to occur at $\rs\approx1.9$. The corresponding transition pressure
would be $2.3Mbar$.\par
However, as $\rs$ increases, the ion-electron coupling becomes stronger,
and the non-linear response of the electron component to the ``external"
potential field provided by the protons, is expected to
lower the free energy of the plasma phase. To explore the non-linear
regime we have adapted the HNC-DFT formulation of our earlier work
on (unpolarized) metallic H~\cite{hxu} to the spin-polarized case.
Within this formulation, proton-proton and proton-electron correlations
are treated at the HNC level, which is expected to be a good
approximation for the long-range Coulomb interactions, while
the energy of the inhomogeneous electron gas follows from the
density functional ($E=N\,\epsilon_e$):
\be
E\left[\rho(\br)\right]=E_K\rror+E_H\rror+E_X\rror+E_C\rror
\label{eq5}
\ee
where $\ror$ denotes the local electron density, and $E_K$,
$E_H$, $E_X$ and $E_C$ are the kinetic, Hartree, exchange
and correlation contributions. For $E_K$ we adopted the Thomas-Fermi
approximation, corrected by a square gradient term:
\be
E_K\rror=C_K\int[\ror]^{5/3}d\br+\frac{\lg}{8}\int
\frac{|{\mathbf{\nabla}}\rho(\br)|^2}{\ror}\dr
\label{eq6}
\ee
where $C_K=3(6\pi^2)^{2/3}/10\;\,a.u.$, while the choice of $1/9<\lg<1$
will be specified below. The mean field Hartree term is of the usual
form:
\be
E_H\rror=\frac{1}{2}\int\dr\int\drp\;\frac{\Delta\ror\Delta\rho(\brp)}
{\rrp}
\label{eq7}
\ee
where $\Delta\ror=\ror-n$, while:
\be
E_X[\rho(\br)]=C_X\int[\rho(\br)]^{4/3}\,\dr
\label{eq8}
\ee
with $C_X=-3(6/\pi)^{1/3}/4\; \,a.u.$ . The correlation contribution
$E_C[\rho(\br)]$ (within the LDA) can be found in \cite{par}. This functional
yields an explicit form for the electron-electron direct correlation
function $c_{22}(r)$ (henceforth the indices 1 and 2 will refer to
protons and electrons respectively)~\cite{hxu}. The remaining direct
and total correlation functions $c_{11}(r)$, $c_{12}(r)$, $h_{11}(r)$
and $h_{12}(r)$ are calculated by a numerical resolution of the
HNC closure equations and the quantum version of the Ornstein-Zernike
(OZ) relations~\cite{chi}, which form a closed set of coupled
non-linear integral equations for the four functions.\par
Solutions were obtained by a standard iterative procedure along
the isotherm $T=10^4K$ and for density parameters in the range
$0.5\leq\rs\leq2.5$, corresponding to more than one-hundred-fold
compression of the lowest density state ($\rs=2.5$), which would
correspond to $0.17gr/cm^3$. The temperature is roughly equal to
that expected inside Saturn, and comparable to temperatures reached in
shock compression experiments on the NOVA laser facility at
Livermore~\cite{col}. The iterative solutions were first obtained at the
highest densities ($\rs=0.5$ and $1$), where linear response theory
provides reasonably accurate intial input. The prefactor $\lg$
in the square gradient correction to the electron kinetic energy functional
(\ref{eq6}) was adjusted to provide the best match between the HNC-DFT
result for the local radial density of electrons around a proton,
$r^2\,g_{12}(r)=r^2[1+h_{12}(r)]$, and its
linear response prediction, at the highest densities ($\rs=0.5$ and $1$),
where linear response should be most accurate; $g_{12}(r)$ turns out
to be rather sensitive to $\lg$, and the best agreement
is achieved for $\lg=0.18$, which is close to the value $1/5$
frequently advocated in electronic structure calculations for
atoms~\cite{par}.\par
%
%fig1
%
Results for the local radial density
$r^2\,g_{12}(r)$ are shown in Fig.1 for several values of $\rs$.
As expected, electrons pile up increasingly at small $r$ as $\rs$
increases, and a shoulder is seen to develop around $r/a\approx0.4$.
For comparison, the linear response prediction is shown at $\rs=1$,
while at the lowest density ($\rs=2.5$), an estimate of $g_{12}(r)$
in the atomic phase is obtained by adding to the electron density
in a $H$ atom (namely $r^2\rho(r)=r^2\exp(-2r)/\pi\;a.u.$) the
convolution of the latter with the atom-atom pair distribution
function $g(r)$. HNC-DFT results for the proton-proton pair distribution
function are shown in Fig.2 for three densities. As expected,
proton-proton correlations are seen at first to weaken, as
the density decreases, due to enhanced electron screening.
However at the lowest density ($\rs=2.5$), weakly damped
oscillations build up at long distances, which may be indicative of
an incipient instability of the proton-electron plasma. The atom-atom
$g(r)$ at the same density agrees reasonably well with $g_{11}(r)$
up to the first peak, but it does not exhibit the long-range
correlations in the latter. The proton-proton structure factors
$S_{11}(k)$ are plotted in Fig.3 for several values of $\rs$.
A considerable qualitative change is again seen to occur at the
lowest density ($\rs=2.5$), where the main peak is shifted to
larger $k$, while a significant peak builds up at $k=0$.
Such enhanced ``small angle scattering" is reminiscent of the
behaviour observed in simple fluids near a spinodal (subcritical)
instability. In fact we were unable to obtain convergence
of the HNC-DFT integral equations for $\rs>2.5$, which hints at
an instability of the electron-proton plasma at lower densities.
This strongly suggests a transition to the insulating atomic
phase, but the simple density functional used in this work
cannot properly describe the recombination of protons and electrons
into bound (atomic) states~\cite{perr}.\par
%
%fig2

In order to confirm this scenario, the free energy of the plasma
phase should be compared to that of the atomic phase. This is easily
achieved within the high density linear response regime, as shown
earlier. However the calculation of the free energy in the 
non-linear regime appropriate for lower densities ($\rs>1$)
is less straightforward~\cite{hxu}. In fact the present HNC-DFT
formulation provides only one direct link with thermodynamics,
namely via the compressibility relation~\cite{han}:
\be
\lim_{k\to 0}S_{11}(k)=\lim_{k\to 0}S_{12}(k)=n\,k_B T\,\chi_{_T}
\label{eq9}
\ee
where $\chi_{_T}$ denotes the isothermal compressibility of the
plasma. From the calculated values of $\chi_{_T}$, the free energy
of the plasma follows by thermodynamic integration, starting from a
reference state (e.g. $\rs=1$) for which the linear response
estimate is expected to be accurate. The resulting ``compressibility"
free energy curve is plotted in Fig.4. Somewhat unexpectedly it lies
above the linear response prediction. An alternative route to the free
energy is via the virial relation for the pressure; only an approximate
virial expression is known within the present HNC-DFT
formulation~\cite{hxu}, and the resulting ``virial" free energy curve
is also shown in Fig.4. It falls well below the ``compressibility"
free energy curve, thus illustrating the well known thermodynamic 
inconsistency of the HNC closure for Coulombic fluids~\cite{hans}. 
Any reasonable extrapolation of the two free energy curves
would miss the low density limit $-0.5\;a.u.$ by as much as 
twenty per cent.
We suggest instead an estimate of the free energy of the plasma
phase by taking the average of the ``compressibility" and ``virial"
values, despite the lack of foundamental a priori justification for doing
this. A short extrapolation of the resulting curve is likely
to intersect or smoothly join on to the free energy of the atomic
phase just beyond $\rs=2.5$. An intersection would correspond to
a first-order phase transition, reminiscent of the ``plasma phase
transition" of Saumon and Chabrier~\cite{sau}. However,
due to the uncertainty on the thermodynamics of the plasma phase,
a continuous transition cannot be ruled out. The transition
pressure $P$ would be of the order of $0.5Mbar$, well below
the current experimental and theoretical estimates for the
transition of fluid molecular Hydrogen to a conducting state~\cite{wei,sau}.\par
%
%fig4

In summary the structure and thermodynamic results derived from
an HNC-DFT theory of the spin-polarized proton-electron plasma strongly
suggest that this plasma will recombine into an insulating
atomic phase at $\rs\approx2.5$, for a temperature $T=10^4K$. We
are presently exploring the behaviour of the system at lower temperatures.\par   
\pagebreak
%%%%%%%%%%%%%%%%%%%%%%%%%%%%%%%%%%%%%%%%%%%%%%%%%%%%%%%%%%%%%%%%%
%\begin{references}      %                    %
             %                    %
\pagebreak
%\include{hydrogbib}
%%%%%%%%%%%%%%%%%%%%%%%%%%%%%%%%%%%%%%%%%%%%%%%%%%%%%%%%%%%%%%%%%

\pagestyle{empty}
\begin{figure}[ht]
\includegraphics[1mm,90mm][180mm,290mm]{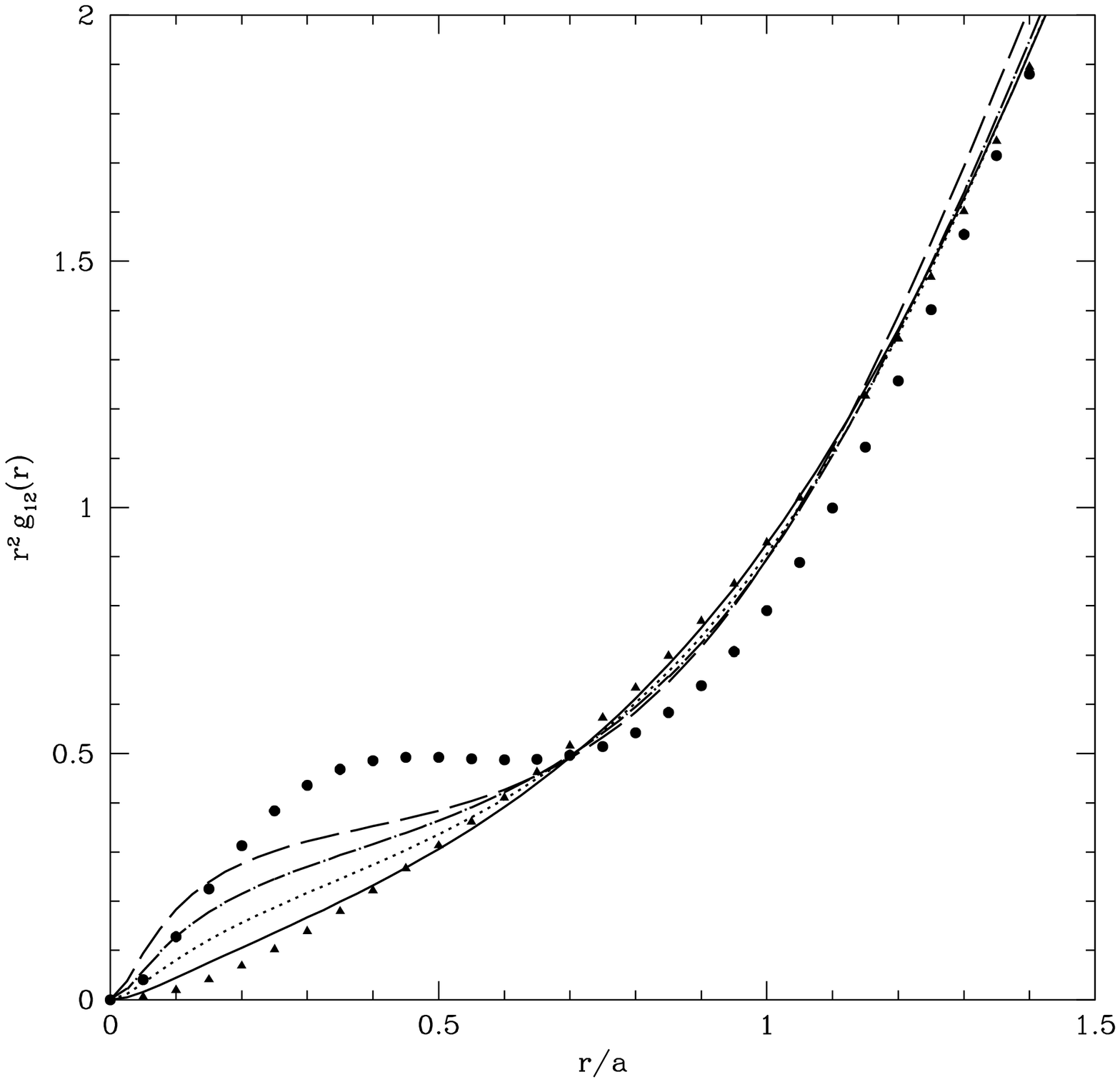}
\caption{ The radial electron density around a proton, $r^2\,g_{12}(r)$,
vs $r/a$, for $\rs=1$(full curve),1.5(dots),2(dash-dotted curve) and
2.5(dashed curve). The triangles are the linear response results at
$\rs=1$; the circles correspond to $r^2\,g_{12}(r)$ in the atomic phase. }
\end{figure}

\begin{figure}[ht]
\includegraphics[1mm,90mm][180mm,290mm]{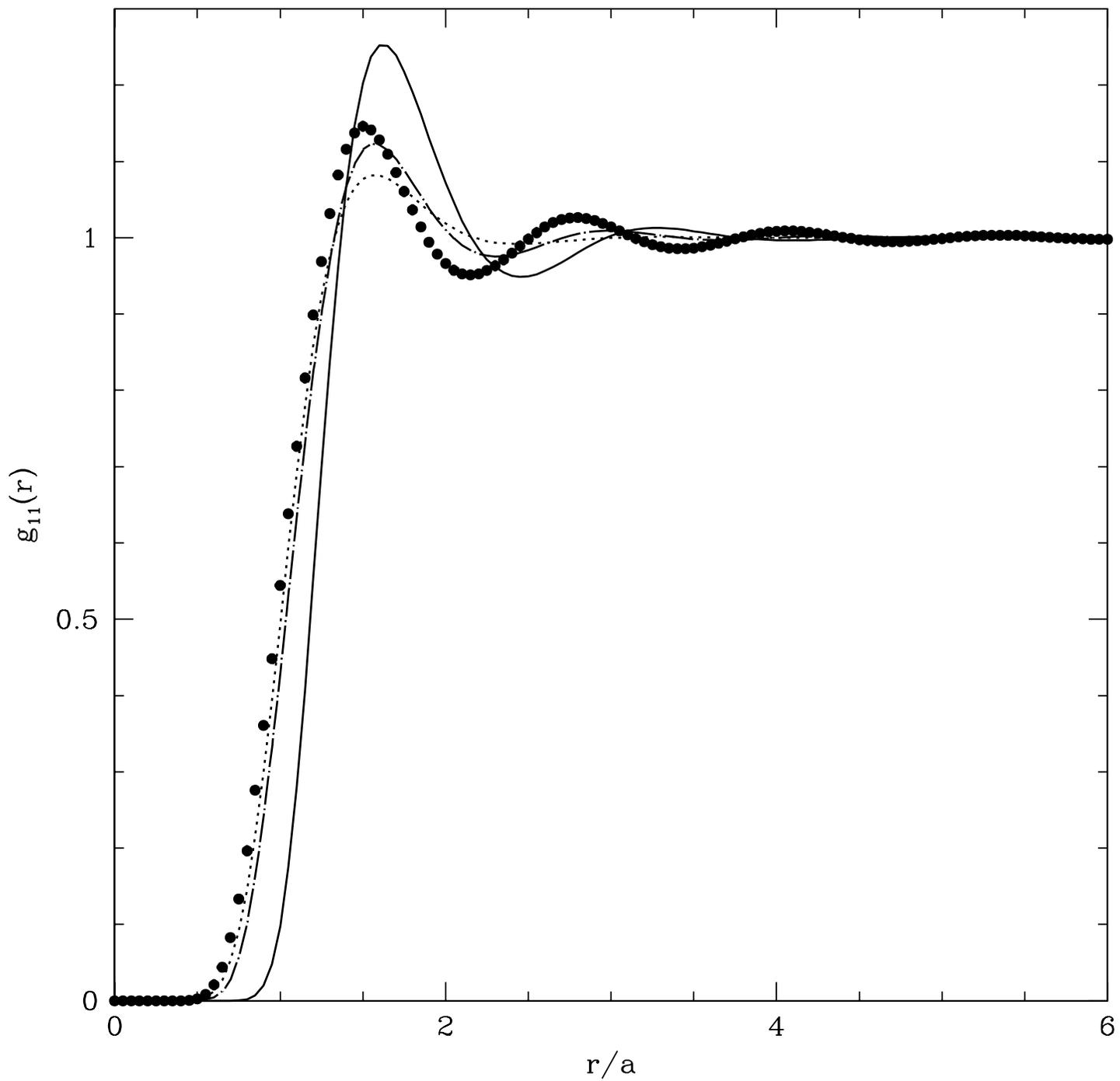}
\caption{The proton-proton pair distribution function $g_{11}(r)$
vs. $r/a$, for $\rs=1$(full curve), 2(dash-dotted curve) and 2.5(circles).
The dotted curve is the atom-atom function $g(r)$ at $\rs=2.5$.}
\end{figure}

\begin{figure}[ht]
\includegraphics[1mm,90mm][180mm,290mm]{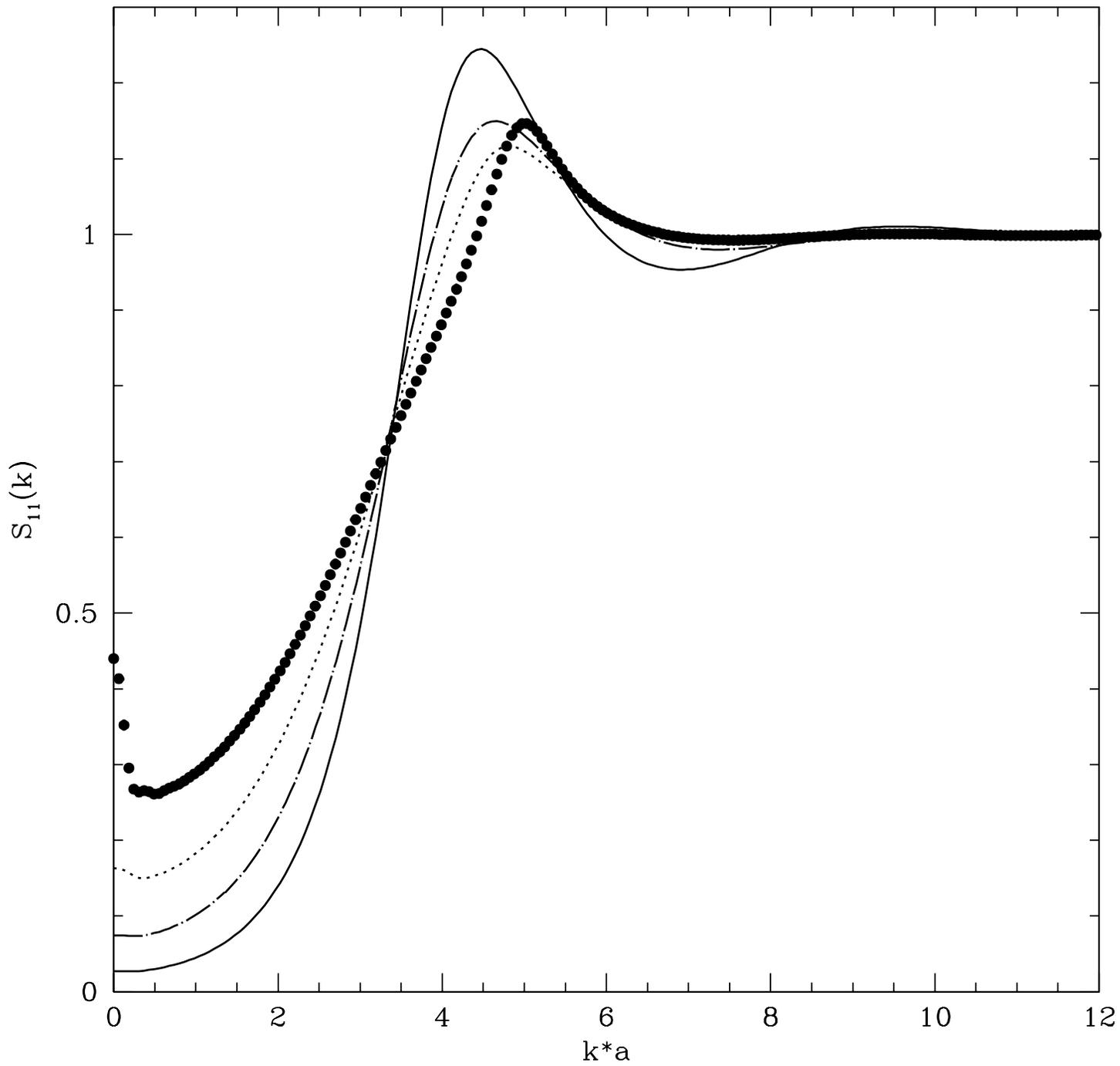}
\caption{The proton-proton structure factor $S_{11}(k)$
vs. $k\,a$, for $\rs=1$(full curve), 1.5(dash-dotted curve),
2(dots) and 2.5(circles).}
\end{figure}

\begin{figure}[ht]
\includegraphics[1mm,90mm][180mm,290mm]{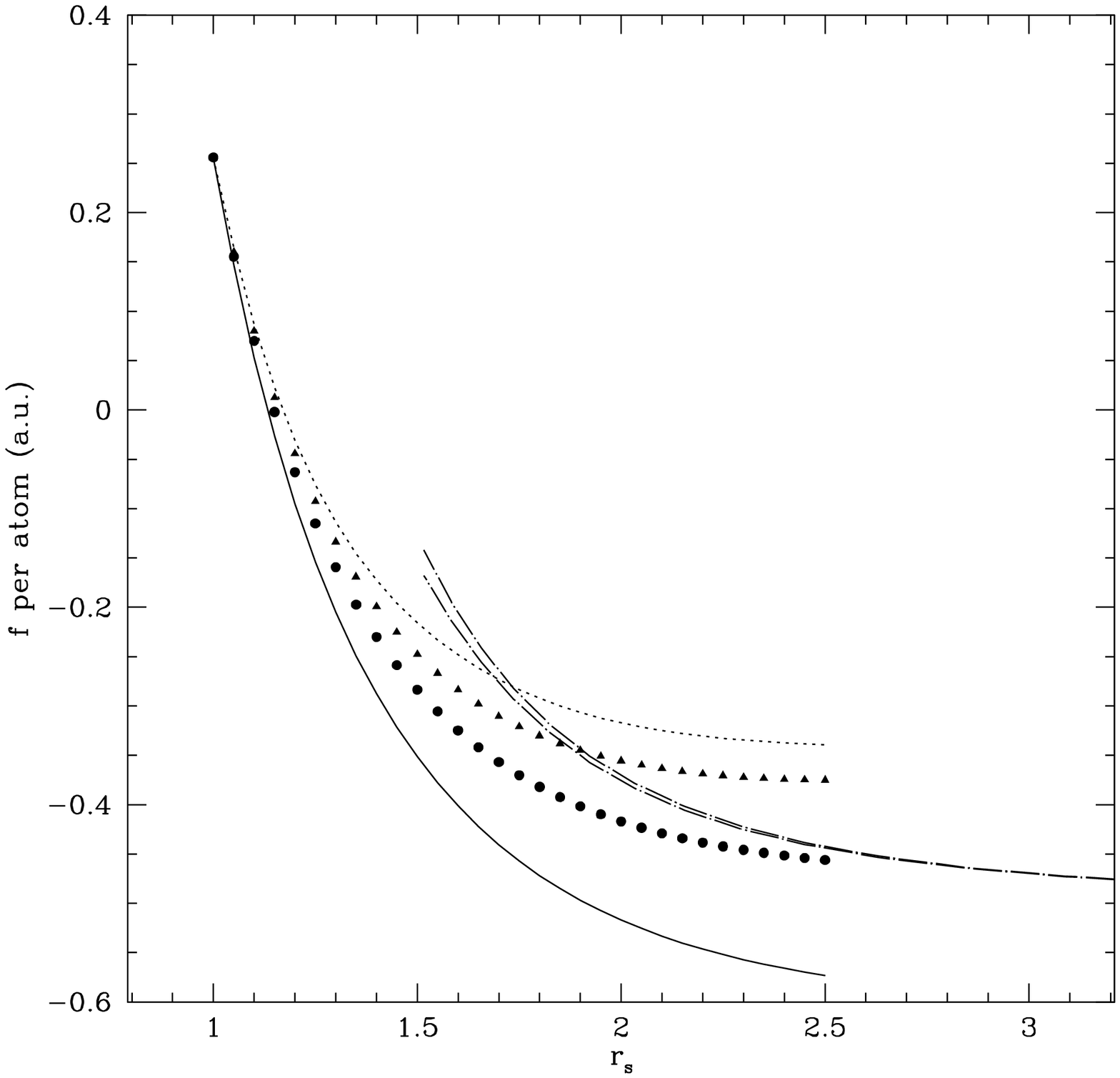}
\caption{Free energy per proton-electron pair (or atom), minus the
ideal proton contribution, versus the density parameter $\rs$.
Full curve: ``virial" free energy of the plasma phase;
dotted curve: ``compressibility" free energy for the plasma phase; circles:
average of the latter two estimates; triangles: linear response result;
dash-dotted curves: ``virial" (upper) and ``compressibility" (lower) 
free energy for the atomic phase.}
\end{figure}


\begin{thebibliography}{999}
%1 
\bibitem{wig}E. Wigner and H.B. Huntington, {\sl J.Chem.Phys.}
{\bf 3}, 764(1935)
%2
\bibitem{che}N.H.Chen, E. Sterer and I.F. Silvera,
{\sl Phys. Rev. Lett.} {\bf 76},1663(1996);
R.J. Hemley, H.K. Mao, A.F. Goncharov, M. Hanfland and V. Struzhkin,
{\sl Phys. Rev. Lett.} {\bf 76},1667(1996)
%3
\bibitem{edw} B. Edwards and N.W. Ashcroft, {\sl Nature},
{\bf 388}, 652(1997)
%4
\bibitem{wei} S.T. Weir, A.C. Mitchell and W.J. Nellis,
{\sl Phys. Rev. Lett.} {\bf 76},1860(1996);
W.J. Nellis, A.A. Louis and N.W. Ashcroft, {\sl Phil. Trans.
R. Soc. London}, {\bf A356}, 119(1998)
%5
\bibitem{col}G.W. Collins et al., {\sl Science}, {\bf 281},
1178(1998)
%6
\bibitem{sau} D. Saumon and G. Chabrier, {\sl Phys. Rev.}
{\bf A46}, 208(1992)
%7
\bibitem{eak} see e.g. C.W. Eaker and C.A. Parr, {\sl J. Chem. Phys.}
{\bf 65}, 5155(1976)
%8
\bibitem{han} see e.g. J.P. Hansen and I.R. McDonald,
{\sl ``Theory of simple liquids"}, 2nd ed. (Academic Press,
London, 1986) 
%9
\bibitem{gal}S. Galam and J.P. Hansen, {\sl Phys. Rev.}{\bf A14}, 816(1976)
%13
\bibitem{par} see e.g. R.G. Parr and W. Yang, {\sl ``Density functional
theory of atoms and molecules"}, (Oxford University Press, New York, 1989)
%10
\bibitem{sla} W.L. Slattery, G.D. Doolen and H.E. DeWitt,
{\sl Phys. Rev.}{\bf A21}, 2087(1980)
%11
\bibitem{hxu} H. Xu and J.P. Hansen, {\sl Phys. Rev.}{\bf E57}, 211(1998)
%12
\bibitem{chi} J. Chihara, {\sl J. Phys. Cond. Matt.} {\bf 3}, 8715 (1991)
%14
\bibitem{perr} F. Perrot,
{\sl J. Phys. Cond. Matter} {\bf 6}, 431(1994)
%14
\bibitem{hans} see e.g. J.P. Hansen, G.M. Torrie and P. Vieillefosse,
{\sl Phys. Rev.}{\bf A16}, 2153(1977)

%\end{references}             %                    %
\end{thebibliography}
\end{document}